\documentclass[10pt,english]{article}
\usepackage{subfigure}
\usepackage{graphicx}
\usepackage{float}
\usepackage[T1]{fontenc}
\usepackage[latin9]{inputenc}
\usepackage[margin=1.25in]{geometry}
\usepackage{float}
\usepackage{amsmath}
\usepackage{amsbsy}
\usepackage{setspace}
\usepackage{amssymb}
\usepackage{esint}
\usepackage{cite}
\newfloat{algorithm}{tbp}{loa}
\floatname{algorithm}{Algorithm}
\usepackage{hyperref}
\usepackage{listings}

\newlength\myindent
\makeatletter

\floatstyle{ruled}
\newfloat{algorithm}{tbp}{loa}
\floatname{algorithm}{Algorithm}

\usepackage{babel}

\begin{document}

\title{Entropically Driven Agents}

\author{M. Andrecut}

\date{February 16, 2025}

\maketitle
{
\centering Unlimited Analytics Inc.

\centering Calgary, Alberta, Canada

\centering mircea.andrecut@gmail.com

} 

\begin{abstract}

Populations of agents often exhibit surprising collective behavior emerging from simple local interactions.  
The common belief is that the agents must posses a certain level of cognitive abilities for such an emerging collective behavior to occur.
However, contrary to this assumption, it is also well known that even noncognitive agents are capable of displaying nontrivial behavior. 
Here we consider an intermediate case, where the agents borrow a little bit from both extremes. 
We assume a population of agents performing random-walk in a bounded environment, on a square lattice. 
The agents can sense their immediate neighborhood, and they will attempt to move into a randomly selected empty site, by avoiding collisions.
Also, the agents will temporary stop moving when they are in contact with at least two other agents. 
We show that surprisingly, such a rudimentary population of agents undergoes a percolation phase transition and self-organizes in a large polymer like structure, as a consequence of an attractive entropic force emerging from 
their restricted-valence and local spatial arrangement. 


\end{abstract}

\section{Introduction}

The emergence of collective behavior in populations of simple agents is a remarkable example of self-organization arising from simple local interactions. 
The common belief is that the emerging collective behavior is a consequence of a certain level of cognitive abilities of the agents, such that the agents are 
able to predict future events according to some optimality criterion. 
For example, recently it has been suggested that such a connection exists 
between causal path entropy maximization and intelligent behavior \cite{key-1}. Causal path entropy ($S_c$) is a measure of diversity of future options available to an agent, 
and its maximization helps driving the agent away from various constraints \cite{key-2}, \cite{key-3}. In this context, the causal path entropy maximization over a time horizon $\tau$, at temperature $T$, results in 
a causal entropic force $\mathbf{F}_c$ \cite{key-4}, \cite{key-5}, steering the agent towards macrostates $\mathbf{X}$ of larger causal entropy:
\begin{equation}
\mathbf{F}_c(\mathbf{X}_0,\tau) = T \nabla_\mathbf{X} S_c (\mathbf{X},\tau).
\end{equation}

The causal path entropy of a macrostate $\mathbf{X}$ is defined as the path integral:
\begin{equation}	
S_c (\mathbf{X},\tau) = -k_B \int_{\mathbf{x}(t)} \text{Pr}(\mathbf{x}(t) \mid \mathbf{x}(0)) \ln \text{Pr}(\mathbf{x}(t) \mid \mathbf{x}(0)) \mathrm{D} \mathbf{x}(t)
\end{equation}
where $k_B$ is Boltzmann's constant, and $\text{Pr}(\mathbf{x}(t) \mid \mathbf{x}(0))$ is the conditional probability of occurrence of path $\mathbf{x}(t)$ starting from $\mathbf{x}_0$:
\begin{equation}
\text{Pr}(\mathbf{x}(t) \mid \mathbf{x}(0)) = \int_{\xi_{\tau}}  \text{Pr}(\mathbf{x}(t),\xi_{\tau} \mid \mathbf{x}(0)) \mathrm{D} \xi_{\tau}
\end{equation}
which is determined by all possible evolutions of the system $\xi_{\tau}$ during the time interval $\tau$.
Thus, such a cognitive agent can build a cognitive map of the environment by estimating the causal path entropy $S_c$ at the current location, and it can navigate by simply following the gradient of $S_c$. 
Following this approach it has been shown that self-organized spatial patterns emerge in populations of cognitive agents who maximize their causal path entropy \cite{key-1}.  

Contrary to the cognitive requirement justification, simple noncognitive examples of emergent collective behavior are provided by 
hard-sphere crystallization \cite{key-6}, and nematic ordering in elongated hard particles \cite{key-7}.
Other examples are the patchy colloids, characterized by the absence of explicit bonding forces, who can self-assemble into interesting structures by exploiting entropic interactions \cite{key-8}.  
An interesting case is that of limited-valence colloids, who have the ability to selectively bind to a controlled number of neighbors, forming larger structures \cite{key-9}. 
In all these cases, the entropy maximization leads to the emergence of an attractive force with the directionality depending on the particle shape and crowding, favoring their self-assembly in ordered structures \cite{key-10}. 
These directional entropic forces are not intrinsic to the particles but instead emerge from their collective behavior. A simplified explanation can be formulated 
assuming that the self-assembly process is driven by the change in the free energy 
of the system that is not isolated, but can exchange energy with its surroundings:
\begin{equation}
F = E - TS,
\end{equation}
Here, $E$ is the internal energy, $T$ the temperature, and $S$ the entropy. 
Assuming that the temperature is constant, in an ordering phase transition the system can lower its free energy either by increasing the entropy, or by decreasing the internal energy \cite{key-11}. 
Therefore, the entropy increase can play a significant role in the self-assembly processes, which is in contradiction to the traditional belief that an increase in entropy immediately translates in higher disorder. 

Inspired by the above examples, in this paper we show that the self-organized collective behavior of the agents can also be considered in the context of locally interacting Markov processes \cite{key-12}. 
The agents correspond to a set of identical random walk particles for which the next configuration state depends only on the current one. 
In such a setting, the local dynamics of the particles depends only on the occupation state of their neighborhood, 
and it must respect the exclusion rule \cite{key-13}, such that at most one particle is allowed to occupy each site of the lattice. 
During this process the initial density of the particles is conserved, and we are interested in the asymptotic diffusion and the transport properties of their spatial distribution.

More specifically, the model consists of a population of agents performing random-walk in a bounded environment, on a square lattice. 
We assume that the agents can sense their immediate neighborhood, and each of them will attempt to move into a randomly selected unoccupied neighboring site, by avoiding collisions. 
We also impose a constraint on the dynamics, such that an agent will temporary stop moving when it is in contact with at least two other agents. 
We show that surprisingly, such a rudimentary population of agents undergoes a percolation phase transition and self-organizes in a large polymer like structure, as a consequence of an attractive entropic force emerging from 
their restricted-valence and local spatial arrangement. 

\section{The stochastic model}

Our stochastic model is inspired by the exclusion process, which is one of the most studied interacting particle systems \cite{key-12},\cite{key-13}. 
The exclusion model consists of a $d-$dimensional lattice where each site $i$ at time $t$ can be in two states: $x_i(t)=1$ if the site is occupied, and $x_i(t)=0$ if the site is empty (free). 
At each time $t$ a site $i$ is randomly chosen, if the site $i$ is occupied then another site $j$ is randomly chosen, and the particle at site $i$ attempts to jump to site $j$. If the site $j$ is empty then the 
particle jumps, otherwise the particle remains at site $i$ (hence the "exclusion process" name), such that the initial density of the particles is conserved. 
Two types of systems are usually described in the context of exclusion processes, and they are typically studied at long asymptotic times, called the hydrodynamic limit. 
These systems have been extensively studied in the literature, suggesting that such local interactions have also global effects.  
For example, depending on the symmetry of the jumps, the exclusion process is related to either the diffusion (heat) equation (symmetric jumps), or the Burgers equation (asymmetric jumps) \cite{key-14}. 

In contrast to the general exclusion process, where the jumps can have any size, here we focus on a more restrictive model, where the lattice is 2-dimensional, and the jumps are limited to 
random moves in the empty sites of the local neighborhood of the agents. Therefore, this limit case corresponds to a population of random walk agents on a 2d lattice. 
Also, in order to introduce a local interaction we assume that the agents cannot move when they are in contact with at least two other agents. Finally, we are interested in the stationary distribution of the agents as a 
a function of their density. 

In a more formal description, we consider a 2-dimensional square lattice $x$ of size $L\times L$. Initially the agents are randomly distributed on the lattice with the probability $p \in (0,1)$.
The agents perform asynchronous avoiding random-walk, and the order of execution is established at the beginning of each round by randomly shuffling the list of agents.
Thus, a simulation round consists of $N$ asynchronous moves, where $N=pL^2$ is the number of agents. 
As mentioned in the introduction, the agents can sense their immediate neighborhood, and at each step of the simulation round, the next agent in the list attempts to move into a randomly chosen unoccupied neighboring site, by avoiding collisions.
However, an agent will temporary stop moving when it is in contact with at least two other agents.
In order to model a finite domain we use reflecting boundary conditions on the lattice. 
Also, the asynchronous dynamics is required in order avoid possible racing conditions, when multiple agents are attempting to move into the same site, at the same time. 
The simulation algorithm is given below:

(1) Set up the empty lattice (array) $x$:
\begin{equation}
x_{ij} \leftarrow 0, \quad 0\leq i,j \leq L-1.
\end{equation}

(2) Compute the neighborhood $\mathcal{N}_{ij} = \{f(i,j) \mid 0\leq i,j \leq L-1\}$ of each site $(i,j)$:
\begin{align}
f(i,j) &\leftarrow 
  \begin{cases}
   (i+1,j)  & \text{if } i<L-1, \; 0 \leq j \leq L-1 \\
   (i-1,j)  & \text{if } i>0, \; 0 \leq j \leq L-1 \\
   (i,j+1)  & \text{if } j<L-1, \; 0 \leq i \leq L-1 \\
   (i,j-1)  & \text{if } j>0, \; 0 \leq i \leq L-1
  \end{cases}.
\end{align}

(3) Populate the lattice with agents by setting a site to 1 with probability $p$:
\begin{equation}
x_{ij} \leftarrow 
  \begin{cases}
   1  & \text{if } r \leq p \\
   0 & \text{otherwise }
  \end{cases}, \quad 0\leq i,j \leq L-1,
\end{equation}
where $r\in (0,1)$ is randomly generated. 

(4) Determine the initial list of occupied sites: 
\begin{equation}
\mathcal{L} \leftarrow [(i,j) \mid x_{ij}=1, 0\leq i,j \leq L-1].
\end{equation}

(5) Randomly shuffle the list of agents $\mathcal{L}$, and set the movement detection variable $\mathcal{M}$ to False:
\begin{align}
\mathcal{L} &\leftarrow \text{random\_shuffle}(\mathcal{L})\\
\mathcal{M} &\leftarrow \text{False}
\end{align}

(6) For each occupied site $(i,j)\in \mathcal{L}$ find the set of free neighboring sites:
\begin{equation}
\mathcal{V}_{ij} \leftarrow \{(n,m) \mid (n,m)\in \mathcal{N}_{ij}, x_{nm}=0, x_{ij}=1\}.
\end{equation}
If $\vert\mathcal{V}_{ij}\vert > 2$ then attempt to move the agent into a randomly chosen free neighboring site, and 
if successful set the movement detection variable to True:
\begin{align}
\text{if }\vert\mathcal{V}_{ij}\vert &> 2: \\
(n,m) &\leftarrow \text{random\_choice}(\mathcal{V}_{ij}) \\
x_{n,m} &\leftarrow 1\\
x_{i,j} &\leftarrow 0\\
\mathcal{M} &\leftarrow \text{True}
\end{align}

(7) If $\mathcal{M} = \text{True}$ then Go To (5).

(8) Return $x$.

\section{Site percolation}

Let us first consider the static case of random site occupation of the finite $L \times L$ lattice. 
Therefore, we assume that the agents are frozen after they are randomly set on the lattice. 
This is the case of the standard site percolation problem in a finite square lattice, where each site is occupied by an agent with probability $p$, or it remains empty with probability $1-p$ \cite{key-15}. 
From the physical point of view, this model is analogous to a random porous medium where each site is filled with a probability $p$. 
It is well known that the occupied sites form clusters. By cluster we mean a set of nearest neighboring occupied sites.
A finite lattice is said to have percolated if a spanning cluster exists, such that it connects its boundaries in a given direction. Here we consider the left-right percolation direction, and 
the top-down choice should be equivalent, since we assume that the lattice is square.

The main question is therefore: what is the minimum probability $p_c$ for the percolation to occur? The probability $p_c$ is also called the percolation threshold. 
An example is shown in Figure 1, and one can see that there is a spanning cluster (yellow) for $p\simeq 0.6$ (Fig. 1(b)).  

The quantity of interest is therefore the minimum probability $\Pi(p,L)$ such that a spanning cluster will emerge, as a function of site occupancy $p$ and the size of the lattice $L$. 
For the standard site percolation problem, the $\Pi(p,L)$ dependence on the site occupation probability and the lattice size $L$ is shown in Figure 2. 
One can see that $\Pi$ exhibits a sharp phase transition for $p\rightarrow p_c \simeq 0.593$, and for larger $L$ values this transition approaches a Heaviside function:
\begin{equation}
\lim_{p\rightarrow p_c} \Pi(p,L) = H(p,p_c) = 
  \begin{cases}
  	1 & \text{if } p>p_c\\
  	0 & \text{otherwise}
  \end{cases}.
\end{equation}
Therefore, the derivative of $\Pi(p,L)$ approaches a Dirac delta distribution when $p\rightarrow p_c$:
\begin{equation}
\lim_{p\rightarrow p_c} \frac{d}{dp} \Pi(p,L) = \delta(p_c),
\end{equation}
which can be used to estimate numerically the critical point $p_c\simeq 0.593$.

\begin{figure}[!ht]
\centering \includegraphics[width=15.2cm]{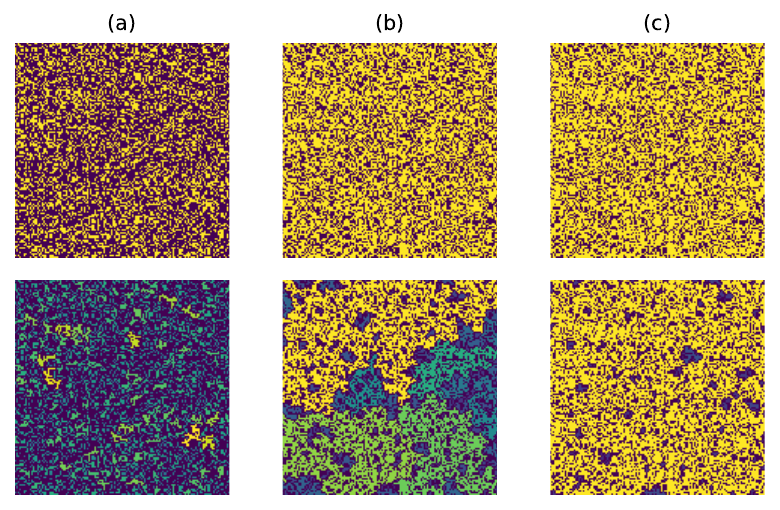}
\caption{Standard site percolation in a square lattice with $L=128$ as a function of the occupation probability: (a) $p=0.35$; (b) $p=0.593$ (percolation cluster); (d) $p=0.65$. 
The top row shows the distribution of particles with the occupation probability $p$, and the bottom row shows the corresponding distribution of clusters. 
}
\end{figure}

\begin{figure}[!ht]
\centering \includegraphics[width=7.55cm]{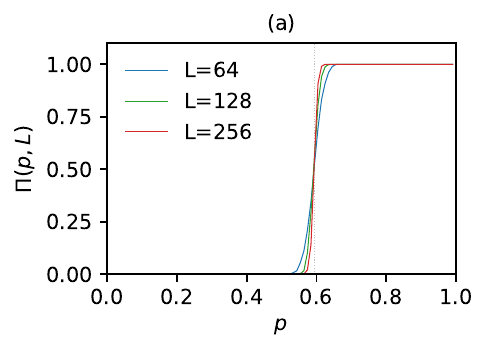}
\centering \includegraphics[width=7.55cm]{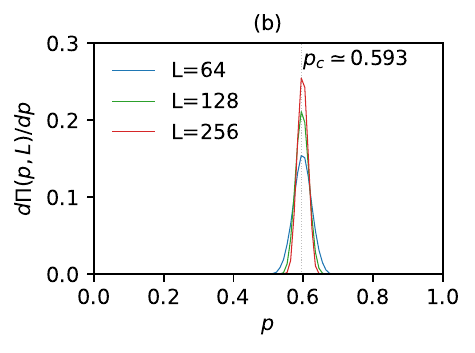}
\caption{Standard percolation: (a) spanning cluster probability $\Pi(p,L)$ as function of $p$ and $L$; (b) the phase transition for $p\rightarrow p_c$.}
\end{figure}

Let us now allow the agents to perform the avoiding random-walk according to the model described in the previous section. 
An example is shown in Figure 3. In the first row we have the initial random distribution of the agents, and in the second row we have the final 
distribution of agents for the same site occupation probability. 
We can see that in this case the agents cluster together in long polymer like chains, and the percolation takes place at a much lower site occupation probability. 
The dependence of the spanning cluster probability $\Pi(p,L)$, of the site occupation probability and the lattice size $L$, is shown in Figure 4. In this case the percolation transition occurs at $p_c \simeq 0.46$. 

\begin{figure}[!ht]
\centering \includegraphics[width=15.2cm]{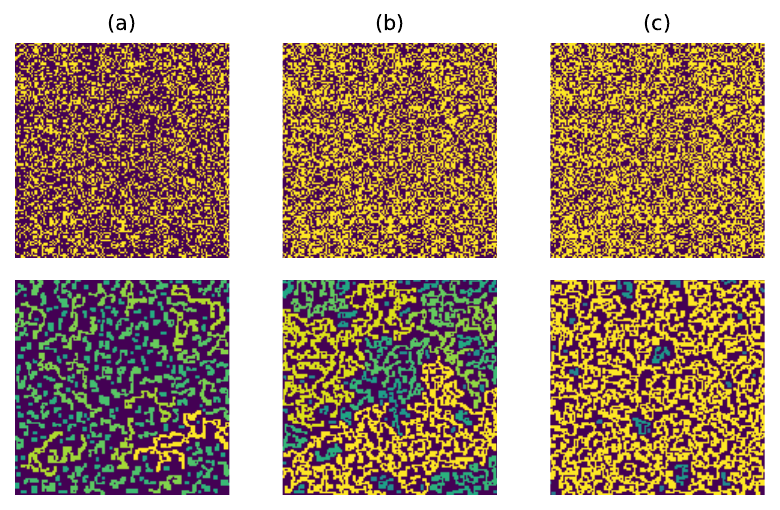}
\caption{Percolation of the random-walk agents in a square lattice with $L=128$ as a function of the site occupation probability: (a) $p=0.35$; (b) $p=0.46$ (percolation cluster); (d) $p=0.5$. 
The top row shows the final distribution of particles with the occupation probability $p$, and the bottom row shows the corresponding distribution of clusters. }
\end{figure}

\begin{figure}[!ht]
\centering \includegraphics[width=7.55cm]{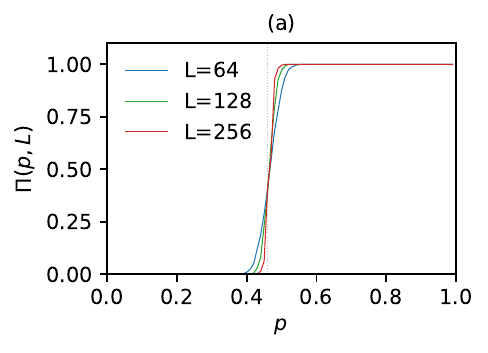}
\centering \includegraphics[width=7.55cm]{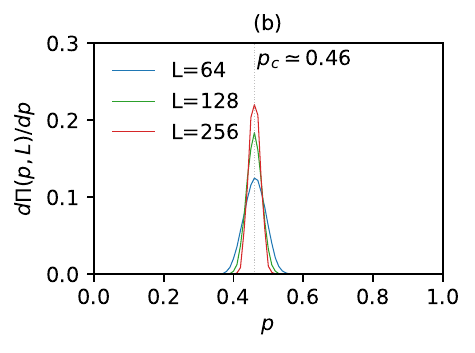}
\caption{Percolation of the random-walk agents: (a) $\Pi(p,L)$ as function of $p$ and $L$; (b) the phase transition for $p\rightarrow p_c$.}
\end{figure}

\begin{figure}[!ht]
\centering \includegraphics[width=7.55cm]{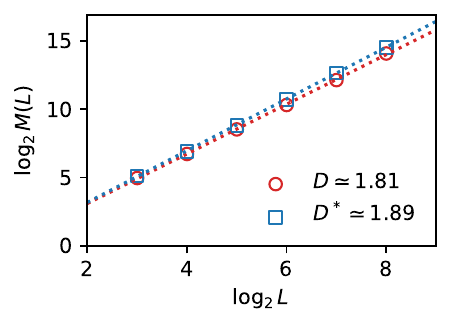}
\caption{The spanning cluster mass as a function of the lattice size: $\log_2 M(L) \propto D\log_2 L$.}
\end{figure}

We notice that at the critical occupation probability $p=p_c$, the spanning cluster has a fractal structure \cite{key-15}, \cite{key-16}. Assuming that $M_s(L)$ denotes the mass of the spanning cluster, then $M_s(L) \propto L^D$, 
where $D$ is the fractal dimension. For a two-dimensional lattice the Euclidean dimension is $d=2$, and the fractal dimension is expected to be smaller $D<d$ when $p \simeq p_c$. However, for $p > p_c$  
the fractal dimension $D$ is expected to be equal to the Euclidean dimension. 
The fractal dimension can be estimated by measuring the mass $M_s(L)$ of the spanning cluster as a function of $L$. 

In Figure 5, we show the results in a double-logarithmic plot, $\log_2 M(L) \propto D \log_2 L$, from where we have estimated $D\simeq 1.81$ for the random-walk agents percolation, 
which a bit smaller than $D^*=91/48\simeq 1.89$ obtained in the case of standard $d=2$ percolation. 

The transport properties of the critical spanning cluster can be estimated by either 
considering the problem of incompressible fluid flow, or equivalently the problem of electrical current flow through random porous media \cite{key-17}, \cite{key-18}, \cite{key-19}. 

The incompressible fluid flow is described by Darcy's law:
\begin{equation}
\phi = \frac{kA}{\eta}\frac{\Delta p}{L},
\end{equation}
where $\phi$ is the fluid volume flowing with viscosity $\eta$, through a cross-sectional area $A$ of a sample with permeability $k$, 
and $\Delta p$ is the pressure drop across the sample length $L$. 
If the sample is $d-$dimensional, then $A\simeq L^{d-1}$, and therefore we have:

\begin{equation}
\phi = L^{d-2} \frac{k}{\eta}\Delta p,
\end{equation}

In the case of the electrical current flow, we notice that the conductance $\gamma$ of a $L^d$ homogeneous material sample with conductivity $\sigma$ is:
\begin{equation}
\gamma = L^{d-2} \sigma, 
\end{equation}
and therefore, in the 2-dimensional case ($d=2$), the conductance and the conductivity have the same value. 
The conductance is an extrinsic property describing how easily the current flows through the material, 
while the conductivity is the intrinsic property, describing the material's inherent ability to conduct electricity. 
Here we use the electrical current approach, since we can avoid the requirement of the material constants $k$ and $\eta$. 

If we apply a voltage $V$ across the critical cluster, then we can compute the current flow through the sample:
\begin{equation}
I = \sigma V,
\end{equation}
where $\sigma$ is the conductivity of the cluster.
Since all the sites in the sample are identical, the conductivity between two adjacent sites $a\equiv (i,j)$ and $b\equiv (n,m)$, with $b\in \mathcal{N}_{a}$ (or $(n,m)\in \mathcal{N}_{ij}$), can be simply defined as following:
\begin{equation}
\sigma_{a,b} = 
  \begin{cases}
  1  & \text{if } x_a = x_b = 1 \\
  0  & \text{otherwise }  
  \end{cases}.
\end{equation}

\begin{figure}[!ht]
\centering \includegraphics[width=15.2cm]{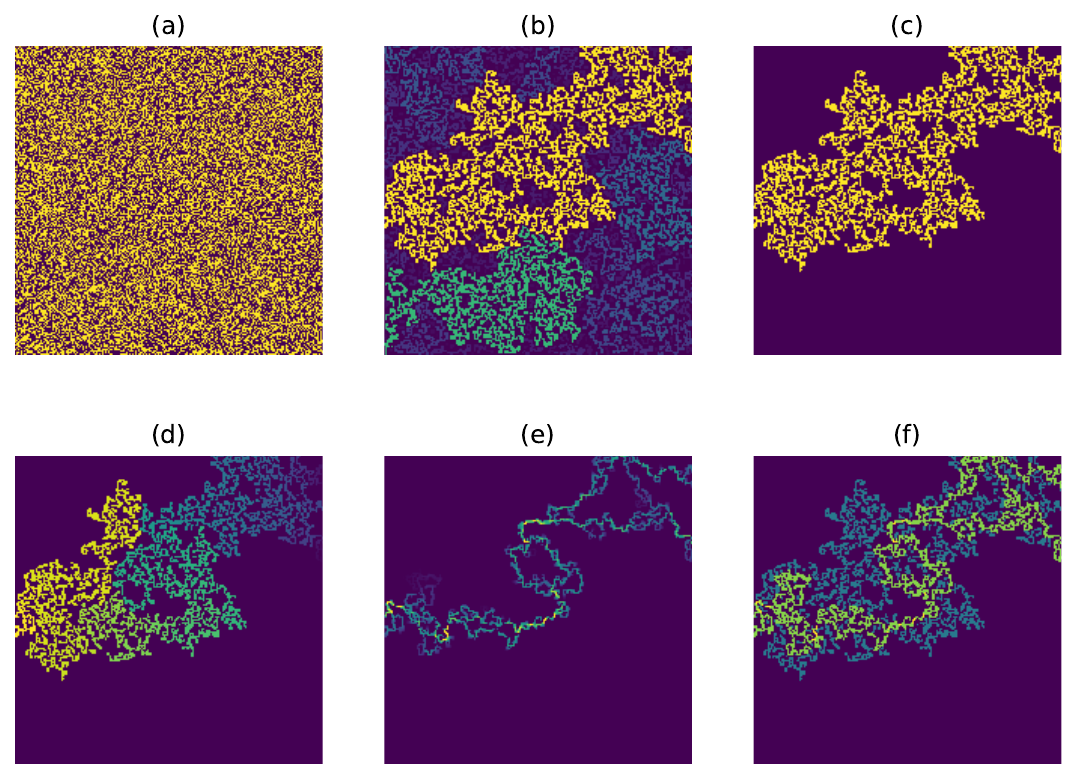}
\caption{Spanning cluster properties: (a) initial condition; (b) percolation; (c) spanning cluster; (d) potential values; (e) current; (f) backbone and dangling ends.}
\end{figure}

Kirchhoff's rule for the currents requires that the sum of the currents flowing from site $a$ into all its neighboring sites $b\in \mathcal{N}_{a}$ must be zero:
\begin{equation}
\sum_b I_{ab} = 0, \quad \sum_b \sigma_{ab} (V_a-V_b) =0.
\end{equation}
In addition, we also have the boundary conditions:
\begin{equation}
V_a = 
  \begin{cases}
  V  & \text{if } a = (0,j), j=0,1,...,L-1 \\
  0  & \text{if } a = (L-1,j), j=0,1,...,L-1 
  \end{cases}.
\end{equation}
Thus, we obtain a system of linear equations from which we can find the potential of each site $V_a$ (here we also assume that $V=1$).
Using the computed potentials and the conductivities, we can find the currents between the neighboring sites:
\begin{equation}
I_{ab}=\sigma_{ab}(V_a-V_b),
\end{equation} 
and subsequently we can calculate the "backbone", which is the section of the critical spanning cluster effectively transporting the current. 
The rest of the cluster corresponds to the "dangling ends", which in principle can be removed, since only the "backbone" contributes to the conductivity. 
Therefore, the scaling exponent of the backbone fractal should be smaller than the scaling exponent $D$ for the spanning cluster. 
The numerical estimation of the backbone's scaling exponent is computationally much harder in this case, because first it requires the random 
walk process to end, which can take a relatively longer time. 
In Figure 6 we give an example of such a calculation for a spanning cluster formed by the random-walk agents. 

\section{Discussion}

It is quite unexpected to see the agents self-organize in such long polymer like chains, without any explicit long range attractive interaction among them. 
Another interesting and counterintuitive phenomenon is that despite of their repulsive (avoiding) random-walk, and independently of their density, 
all the agents are "attracted" and consequently "absorbed" into growing clusters, self-organizing into an intricate structure that percolates the lattice at a critical concentration. 
The main question is therefore: what is the origin of this "attracting" force acting on the agents?

Typically, order appears in many-body systems due to long range attractive forces overcoming an increasing entropy. 
In our case, in the absence of any long range attractive interactions, 
the main driving force of the self-organization process must be emerging from the entropy gradient. 

The entropy of a random variable is the average uncertainty of variable's potential outcomes. Assuming that the random variable $X$ takes 
discrete values in the set $\mathcal{X}$, and it is distributed according to the probability distribution $p: \mathcal{X} \rightarrow [0,1]$ the entropy of of $X$ is \cite{key-20}:
\begin{equation}
S(X) = - \frac{1}{\log_2|\mathcal{X}|} \sum_{x\in  \mathcal{X}} p(x) \log_2 p(x) \in [0,1].
\end{equation}

Thus, the entropy is always positive and reaches its maximum for a uniform distribution of outcomes. 
Unfortunately, the entropy as a measure of complexity fails to capture the complexity of two dimensional patterns. 
For example, in our case each site can take the binary values $\{0,1\}$, with a probability $1-p$, and respectively $p$. If the random variable we are considering is 
the site occupation, and since the number of particle is conserved, we always have:
\begin{equation}
S = -\sum_{x\in  \{0,1\}} p(x) \log_2 p(x) = -p \log_2 p - (1-p)\log_2 (1-p)  \in [0,1].
\end{equation}
This entropy has a maximum value $S_{max}=1$ for $p=1/2$, and it does not reveal anything about the complex dynamics of the agents. 

A much better option would be to consider the entropy of the clusters.  
Let us assume that for the site occupancy probability $p$ we have $K_p$ clusters with the area (number of particles) $\{\mathcal{A}_k\}_{k=0,...,K_p-1}$, then we can define the probability distribution of the cluster sizes:
\begin{equation}
q_j(p) = \frac{ \mathcal{A}_j }{\sum_{k=0}^{K_p-1}  \mathcal{A}_k }  \in [0,1], \quad j=0,...,K_p-1,
\end{equation}
such that:
\begin{equation}
\sum_{j=0}^{K_p-1} q_j(p) = 1,
\end{equation}
and consequently the entropy of the clusters distribution can be defined as following:
\begin{equation}
S_c(p) = -\frac{1}{\log_2 K_p}\sum_{k=0}^{K_p-1} q_k(p) \log_2 q_k(p) \in [0,1].
\end{equation}
We can average this entropy, $\langle S_c(p,L) \rangle$, over multiple runs for each value of $p\in[0,1]$, and lattice size $L$. The results are shown in Figure 7, and we can see that we obtain a 
critical transition at the same $p_c$ values as before. 
In Figure 7(a) we have the critical transition for the standard percolation problem, while in Figure 7(b) we have the critical transition for the random-walk agents 
(computed for each $p$ after the dynamics stops, and all the agents 
have been absorbed into clusters). We can see that at low values of $p$ we have a high entropy value, 
corresponding to the presence of many small clusters, while at higher values of $p$ the entropy decreases, since the number of clusters decreases, and reaches zero when 
all the agents are absorbed into a single giant spanning cluster. 

\begin{figure}[!ht]
\centering \includegraphics[width=7.55cm]{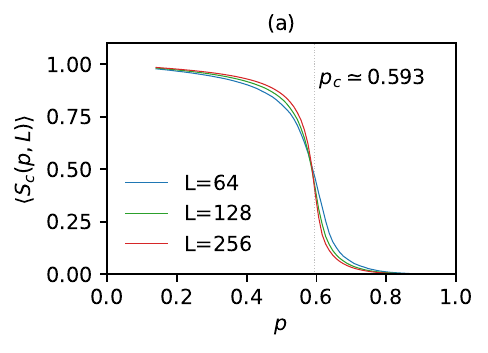}
\centering \includegraphics[width=7.55cm]{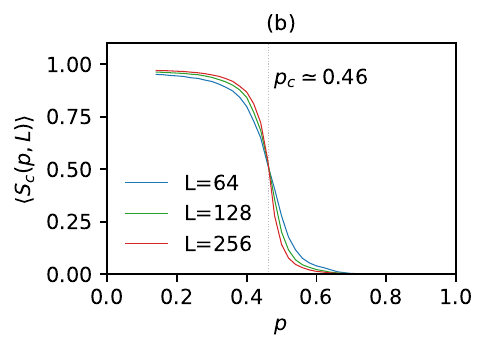}
\caption{The entropy of the clusters distribution $\langle S_c(p,L) \rangle$ as a function of the site occupation probability: (a) standard site percolation model; (b) random-walk agents percolation model.}
\end{figure}

Obviously, neither of these entropy measures can explain the apparent "attractive" force arising between the agents, and overcoming the repulsive random-walk dynamics.  
The first entropy measure (28) is defined at the smallest possible scale (single site), and therefore it fails to capture any collective behavior. Contrary to this, the second entropy measure (31) captures the collective behavior 
and the cluster organization at a global scale, unfortunately it is too coarse and it cannot capture the details of the self-organization process at intermediate scales. 
Therefore, we need an entropy measure that can be defined at the neighborhood level, because this is where the "absorption" process of agents takes place. 

\begin{figure}[!ht]
\centering \includegraphics[width=6 cm]{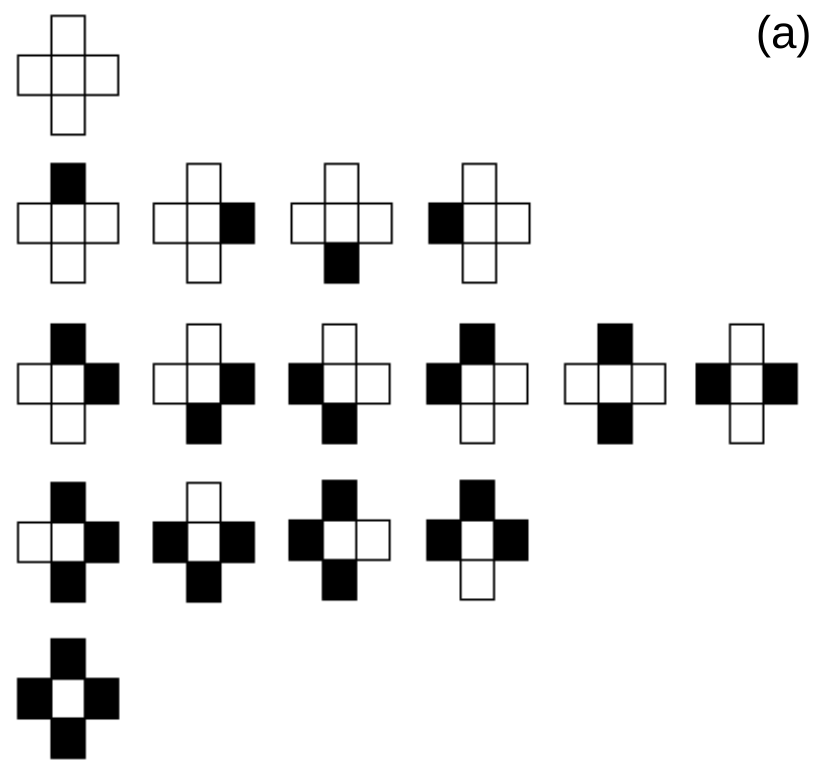}
\centering \includegraphics[width=6 cm]{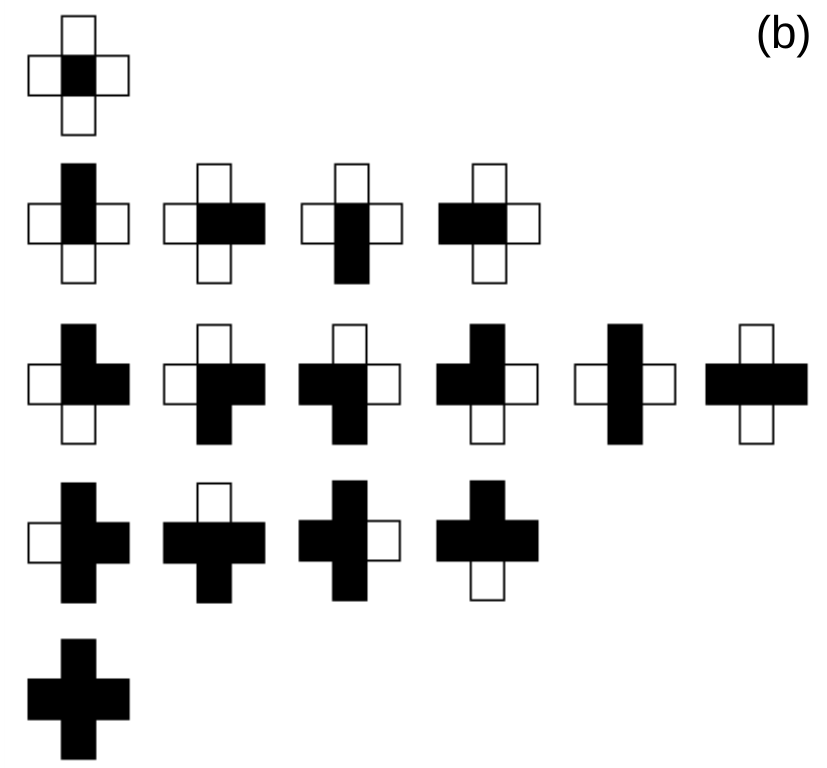}
\caption{The possible configurations of a site neighborhood: (a) empty central site; (b) occupied central site.}
\end{figure}

We notice that in the case of the square lattice considered here, there are $32$ neighborhood configurations, where 16 of them have an empty central site and the other 16 have an occupied central site, as shown in Figure 8.
We are especially interested on the first 16 configurations (Fig. 8(a)), since these are the ones that can accept an agent moving into the empty central site.
Among these configurations, only the last five are temporary "absorbing", by allowing moves into the unoccupied center with at least two neighbors.

We also notice that if we exclude the central site, the number of occupied sites in these neighborhood configurations can be $s\in \{0,1,2,3,4\}$. 
Therefore, since the entropy is an additive quantity, it can be decomposed in two components:
\begin{equation}
S(p) = S_0(p) + S_1(p),
\end{equation}
where $S_0(p)$ is the entropy of the neighborhoods with an unoccupied central site:
\begin{equation}
S_0(p) = -\frac{1}{\log_2 10}\sum_{s=0}^4 p(s | x_{ij}=0) \log_2 p(s | x_{ij}=0) \in [0,1],
\end{equation}
and respectively $S_1(p)$ is the entropy of the neighborhoods with an occupied central site:
\begin{equation}
S_1(p) = -\frac{1}{\log_2 10}\sum_{s=0}^4 p(s | x_{ij}=1) \log_2 p(s | x_{ij}=1) \in [0,1]. 
\end{equation}
Here, $p(s | x_{ij}=k)$, $k\in \{0,1\}$ is the conditional probability that the number of occupied sites in the neighborhood is $s\in \{0,1,2,3,4\}$, given  that the central site is $x_{ij} \in \{0,1\}$. 
We average these components of the entropy, $\langle S_0(p,L) \rangle$ and $\langle S_1(p,L) \rangle$, over multiple runs for each value $p\in [0,1]$, and lattice size $L$, and the results are shown in Figure 9. 

\begin{figure}[!ht]
\centering \includegraphics[width=7.55cm]{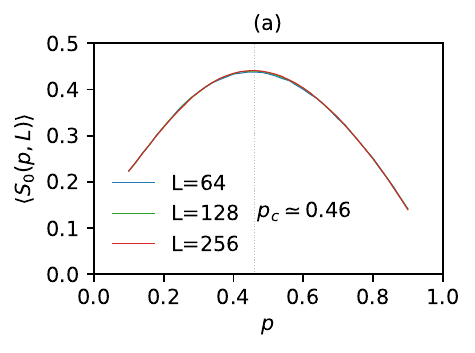}
\centering \includegraphics[width=7.55cm]{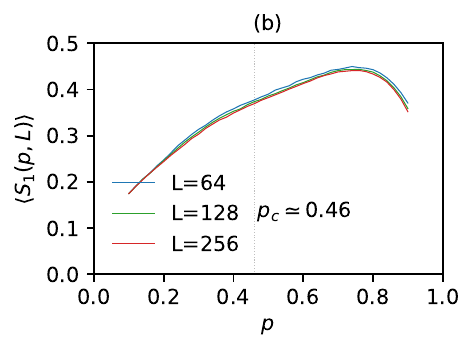}
\caption{The entropy of the neighborhood configurations for the random-walk model: (a) empty central site  $\langle S_0(p,L) \rangle$; (b) occupied central site $\langle S_1(p,L) \rangle$.}
\end{figure}
We can see that the component $\langle S_0(p,L) \rangle$ of the neighborhood entropy measure  $\langle S(p,L) \rangle$  has a maximum at the same critical percolation value $p_c\simeq 0.46$. 
For $p<p_c$ the dynamics of the agents creates an increasing number of "absorbing" sites, while for $p>p_c$ this number starts to decrease because the "absorbing" sites become occupied. 
Also, it is interesting to note that in the case of $\langle S_1(p,L) \rangle$, the critical point $p_c \simeq 0.46$ seems to become an undulation point, which is a point on a curve where the curvature vanishes but it does not change sign. 
The $\langle S_1(p,L) \rangle$ component is increasing until $p\simeq 0.76$, when it is starting to decrease sharply because the distribution of neighborhoods becomes more skewed towards full occupation configurations. 
Thus, our hypothesis that $\langle S_0(p,L) \rangle$ is the entropy contribution responsible for the "attractive" force of the self-organization process seems to be confirmed. 

\section{Conclusion}

The majority of many-body systems interact through long-range attractive forces, overcoming the disordering effects of entropy. 
In contradiction to this assumption, patchy particles and colloidal particles have the ability to self-organize into large structures, without relying on attractive long-range interactions. 
These particles exploit the directional entropic forces emerging from their geometric features, or their restrictive valence, facilitating local dense packing. Therefore, their self-assembly in ordered structures 
it is expected to become more important as the system becomes crowded. 
In this paper we have considered such an example of self-organizing agents, without relying on any long-range attractive interactions. 
The model consists of a simple population of agents who continuously perform random-walk in a bounded environment, on a square lattice. 
The agents can sense their immediate neighborhood, and each of them will attempt to move into a randomly selected empty site, by avoiding collisions. 
The agents are characterized by limited-valence, and an agent will temporary stop moving when it is in contact with at least two other agents.
We have shown that unexpectedly, the agent population undergoes a percolation phase transition and self-organizes in a large polymer chain like structure, as a consequence of an attractive entropic force arising from 
their limited-valence and local spatial arrangement. Also we have determined numerically the percolation threshold and the fractal dimension of the critical spanning cluster. 
The details of the simulation, and the parameters, are provided in the Appendix, together with a Python program for simulating the proposed model. 

\section*{Appendix}

The computations were performed in Python, averaging over $T=10^2$ simulations for each $p\in [0,1]$, using a step $\Delta p = 10^{-2}$. 
Due to the substantial length of the complete code, below we give just a minimal implementation of the animation section of the 
random walk agents, and the clustering of their final configuration (\url{https://github.com/mandrecut/entropically_driven_agents}). 

\begin{small}
\begin{verbatim}
import numpy as np
from scipy.ndimage import label
from scipy.ndimage import sum_labels
import matplotlib.pyplot as plt
import matplotlib.animation as animation

def neighbors(L):
    v = [(-1,0),(1,0),(0,-1),(0,1)]
    w = np.zeros((L,L),dtype="object")
    for n in range(L):
        for m in range(L):
            w[n,m] = [x for x in v if n+x[0]>=0 and n+x[0]<L and m+x[1]>=0 and m+x[1]<L]
    return w
          
if __name__ == "__main__":
    p = 0.465 # occupation probability
    L = 128 # lattice size    
    T = 1000 # max number of time steps
    a = (np.random.rand(L,L)<p).astype("int") # initial populated lattice
    g = neighbors(L) #list of neighbors for each site
    # animation, adjust interval for increasing the speed                          
    fig, ax = plt.subplots(figsize=(5,5))
    ax.axis('off')
    ims,im = [],ax.imshow(a,animated=True)
    ims.append([im])        
    x = np.array([i for i in range(L)])
    y = np.array([j for j in range(L)])    
    for t in range(T):
        np.random.shuffle(x)
        np.random.shuffle(y)
        flag = True        
        for n in range(L):
            for m in range(L):
                if a[x[n],y[m]] == 1:
                    w = [a[x[n]+c[0],y[m]+c[1]] for c in g[x[n],y[m]]]
                    if np.sum(w) < 2:                        
                        u = [g[x[n],y[m]][i] for i in range(len(w)) if w[i] == 0]
                        if len(u) > 0:
                            (q,r) = u[np.random.randint(len(u))]
                            a[x[n]+q,y[m]+r] = a[x[n],y[m]]
                            a[x[n],y[m]] = 0
                            flag = False                                                
        im = ax.imshow(a,animated=True)
        ims.append([im])
        if flag:
            break                
    ani = animation.ArtistAnimation(fig,ims,interval=100,blit=True,repeat=False)
    plt.show()       
    # find and display clusters
    fig = plt.figure(figsize=(5,5))
    w,n = label(a)
    area = sum_labels(a,w,index=np.arange(n+1)).astype("int")
    plt.imshow(np.sqrt(area[w]),origin='lower',interpolation='nearest')
    plt.axis("off")    
    plt.tight_layout()          
    plt.show()
\end{verbatim}
\end{small}

\end{document}